\documentclass[12pt]{article}                                             %
\usepackage{amsmath,amssymb,amsfonts,amsbsy}                              %
\usepackage{cite}                                                         %
\usepackage{graphicx}                                                     %
\usepackage{wrapfig}                                                      %
\usepackage{epsfig}                                                       %
\usepackage{bbm} 
\usepackage{bm} 
\usepackage{color}                                                       
\usepackage{dsfont} 
\usepackage{latexsym} 
\usepackage{lscape} 
\usepackage{mathrsfs} 
\usepackage{morefloats} 
\usepackage{floatflt} 
\usepackage{slashed} 
\usepackage{psfrag}                                                       %

\begin{document}
\graphicspath{{azim_asym_pic/}} 

\begin{center}
\textbf{AZIMUTHAL ASYMMETRIES IN PRODUCTION OF\\ CHARGED HADRONS 
BY HIGH ENERGY MUONS\\ ON POLARIZED DEUTERIUM TARGETS}\footnote{ 
Supported by the RFFI grant 08-02-91013 CERN{\_}a}

\vspace{5mm}
I.A. Savin on behalf of the COMPASS collaboration

\vspace{5mm}
\begin{small}
\emph{Joint Institute for Nuclear Research, Dubna} \\
\end{small}
\end{center}

\vspace{0.0mm} 

\begin{abstract}
Search for azimuthal asymmetries  in semi-inclusive production of 
charged hadrons by 160 GeV muons on the longitudinally polarized 
deuterium target, has been performed using the 2002- 2004 COMPASS 
data. The observed asymmetries integrated over the kinematical 
variables do not depend on the azimuthal angle of produced 
hadrons and are consistent with the ratio $g_1^d(x)/f_1^d(x)$. 
The asymmetries are parameterized taking into account possible 
contributions from different parton distribution functions and 
parton fragmentation functions depending on the transverse spin 
of quarks.They can be modulated (either/or/and) with 
$\sin(\phi),\,\sin(2\phi),\,\sin(3\phi)$ and $\cos(\phi)$. The 
$x$-, $z$- and $p_h^T$-dependencies of these amplitudes are 
studied.
\end{abstract}

\vspace{7.2mm}

\paragraph{1. Introduction.}

Although the longitudinal spin structure of nucleons has been 
investigated for more than 20 years and results are very well 
known, the studies of the transverse spin structure of nucleons 
have been started recently. Since the pioneering HERMES 
\cite{[1]} and CLAS \cite{[2]} experiments it is known that the 
signature of the transverse spin effects is an appearance of 
azimuthal asymmetries (AA) of the hadrons produced in 
Semi-Inclusive Deep Inelastic Scattering (SIDIS) of leptons on 
polarized targets. 

These asymmetries  are related with new Parton Distribution 
Functions (PDF) and new polarized Parton Fragmentation Functions 
(PFF), depending on the transverse spin of quarks \cite{[3]}. AAs 
on the transversally polarized targets have been already reported 
by HERMES \cite{[4]} and COMPASS \cite{[14],[5]}, and on the 
longitudinally polarized targets - by HERMES \cite{[6],[7]}. The 
search for the AA using the COMPASS spectrometer \cite{[8]} with 
the longitudinally polarized deuterium target is described below.

\begin{figure}[htbp]
\begin{tabular}{cc}
\includegraphics[width=.4\textwidth]{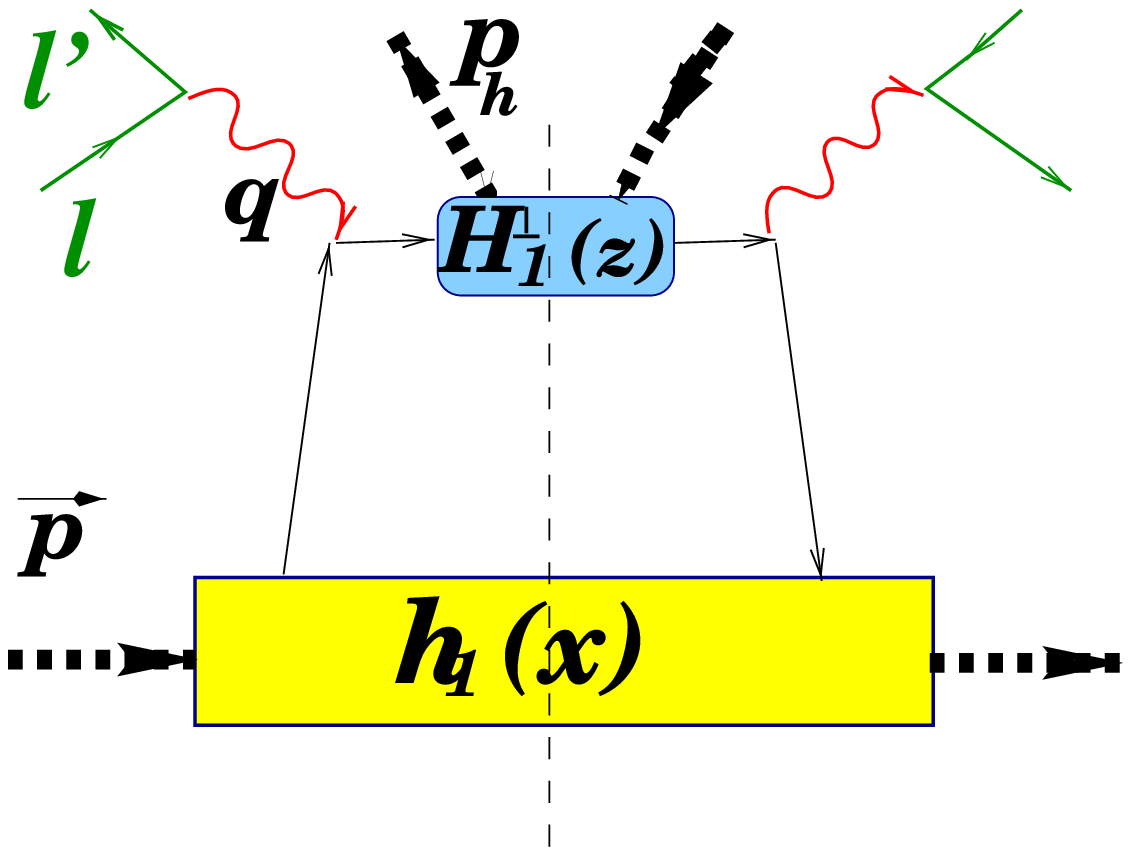}&
\raisebox{48mm}{\rotatebox{-90}{
\includegraphics[width=.24\textwidth]{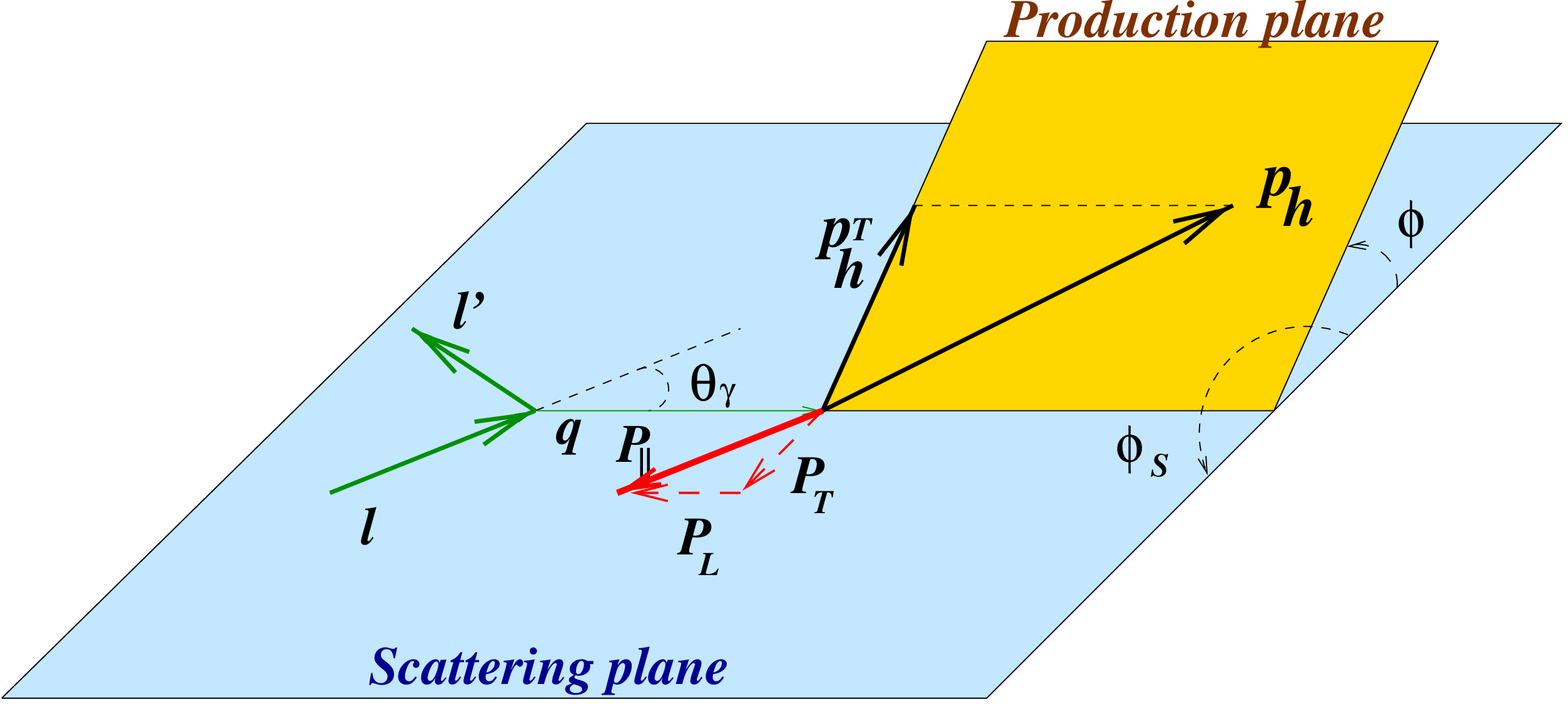}}}\cr
{\bf(a)} & {\bf(b)}
\end{tabular}
\caption{\label{fig1}\footnotesize The squared modulus of the matrix 
element of the SIDIS reaction $\ell+\vec{N}\to\ell'+h+X$ summed 
over X states {\bf(a)} and kinematics of the process {\bf(b)}.}
\end{figure}

In the framework of the parton model of nucleons, the squared 
modulus of the matrix element of the SIDIS is represented by the 
type of the diagram in Fig.\ref{fig1}a, where an example of one 
of the new PDF, transversity, $h_{1}(x)$ and new Collins PFF, 
$H_1^\bot(z)$, is shown. New PDFs and PFFs, due to their chiral 
odd structure, always appear in pairs.

The kinematics of the SIDIS is shown in Fig.\ref{fig1}b, where 
$\ell\, (\ell')$ is the 4-momentum of incident (scattered) 
lepton, $q=\ell-\ell', \quad Q^2 = - q^2, \quad \theta _\gamma $ 
is the angle of the virtual photon momentum $\vec {q}$ with 
respect to the beam, $P_L(P_T )$ is a longitudinal (transversal) 
component of the target polarization, $P_{II}$, with respect to 
the virtual photon momentum in the laboratory frame, $p_h$ is the 
hadron momentum with the transverse component $p_h^T$, $\phi$ is 
the azimuthal angle between the scattering plane and hadron 
production plane, $\phi_S$ is the angle of the target 
polarization vector with respect to the lepton scattering plane 
(for the longitudinal target polarization $\phi_S=0$ or $\pi$. 
For the target polarization $P_{II}$, which is longitudinal with 
respect to the lepton beam, the transverse component is equal to 
$\left|{P_T}\right|=P_{II}\sin(\theta_\gamma)$, where 
$\sin(\theta_\gamma)\approx 2\frac{M}{Q}x\sqrt{1 - y}$, $y= 
\frac{q\cdot p}{p\cdot\ell }$ and $M$ is the nucleon mass. The 
Bjorken variable $x$ and the hadron fractional momentum $z$ are 
defined as $x=Q^2/2p\cdot q, \quad z=p\cdot p_h/p\cdot q$, where 
$p$ is the 4-momentum of the incident nucleon. 

In general, the total cross  section of the SIDIS reaction is a 
linear function of the lepton beam polarization,$P_\mu$, and of the 
target polarization $P_{II}$ or its components:
\begin{equation}
\label{eq1} 
d\sigma=d\sigma_{00}+P_\mu d\sigma_{L0}+P_L\left( 
{d\sigma_{0L} +P_\mu d\sigma_{LL}}\right)+\left|{P_T} 
\right|\left({d\sigma_{0T}+P_\mu d\sigma_{LT}}\right)\,,
\end{equation}
where the first (second) subscript of the partial cross sections 
means the beam (target) polarization.

The asymmetry, a($\phi)$, in the hadron production from the 
longitudinally polarized target (LPT), is defined by the 
expression:
\begin{equation}
\label{eq2} 
a(\phi)=\frac{d\sigma^{\rightarrow\Rightarrow}-d\sigma^{\rightarrow\Leftarrow}} 
{d\sigma^{\rightarrow\Rightarrow}+d\sigma^{\rightarrow\Leftarrow}}  
\propto P_L\left({d\sigma_{0L}+P_\mu d\sigma_{LL}}\right) 
+\left|{P_L}\right|\sin(\theta_\gamma)\left({d\sigma_{0T}+P_\mu 
d\sigma_{LT}}\right)\,.
\end{equation}

Each of the partial cross sections is characterized by the  
specific dependence of the definite convolution of PDF and PFF 
times a function the azimuthal angle of the outgoing hadron. 
Namely, contributions to Eq. (\ref{eq2}) from each quark and 
antiquark flavor, up to the order $(M/Q)$, have the forms: {\small
\begin{eqnarray}
d\sigma_{0L}\!\!\!\!\! &\propto&\!\!\!\!\! \epsilon xh_{1L}^\bot 
(x)\otimes H_1^\bot(z)\sin(2\phi)+\sqrt{2\epsilon(1-\epsilon)}
\frac{M}{Q}x^2\left(h_L (x)\otimes H_1^\bot(z)
+f_L^\bot(x)\otimes D_1(z)\right)\sin(\phi),\nonumber\\
d\sigma_{LL}\!\!\!\!\! 
&\propto&\!\!\!\!\!\sqrt{1-\epsilon^2}xg_{1L}(x)\otimes D_1(z) + 
\sqrt{2\epsilon(1-\epsilon)}\frac{M}{Q}x^2
\left(g_L^\bot(x)\otimes D_1(z)
+e_L(x)\otimes H_1^\bot(z)\right)\cos (\phi),\nonumber\\
d\sigma_{0T}\!\!\!\!\! 
&\propto&\!\!\!\!\!\epsilon\{xh_1(x\!)\!\otimes\! 
H_1^\bot\!(z\!)\sin(\phi\!+\!\phi_S\!)\!+\! 
xh_{1T}^\bot(x\!)\!\otimes\! H_1^\bot\!(z\!)\sin(3\phi\!-\!\phi_S\!) 
\!-\!xf_{1T}^\bot (x)\!\otimes\! D_1\!(z)\sin(\phi\!-\!\phi_S\!)\},\nonumber\\
d\sigma_{LT}\!\!\!\!\! &\propto&\!\!\!\!\!\sqrt{1\!-\!\epsilon^2} 
{xg_{1T}(x\!)\! \otimes\! D_1(z\!)\cos(\phi-\phi_S\!)}\,,\label{eq3} 
\end{eqnarray}}
\noindent where $\otimes$ is a convolution in parton's internal 
transversal momentum, k$_{T}$ , on which PDF and PFF depend, 
$\phi _{s}$=0 for the LPT and 
$\epsilon\approx\frac{2(1-y)}{2-2y+y^2}$. The structure of the 
partial cross sections and physics interpretations of the new 
PDFs and PFFs, entering in a($\phi )$, are given in 
\cite{[9],[10],[11]}.

So, the aim of this  study is to see the AA in the hadron 
production from LTP, as a manifestation of new PDFs and PFFs and 
the $x$, $z$ and $p_h^T$- dependence of the corresponding 
amplitudes.

\paragraph{2. Method of analysis.}

The COMPASS polarized target \cite{[8]} in 2002-2004 years had 
two cells, Up- and Down-stream of the beam, placed in the 2.5 T 
solenoid magnetic field. The target material of the cells 
($^{6}$LiD or NH$_{3})$ can be polarized in opposite directions 
with respect to the beam, for example in the U-cell along to the 
beam (positive polarization) and in D-cell -- opposite to the 
beam (negative polarization) and vice versa. Such a configuration 
can be achieved by means of the microwave field at low 
temperatures at any direction of the solenoid magnetic field 
holding the polarization. Suppose that the above configuration of 
the cell polarizations is realized with the positive (along to 
the beam) solenoid field, then, to avoid possible systematic 
effects in acceptance connected with this field, after some time 
the same configuration of polarizations is realized by means of 
the microwave field with the negative (opposite to the beam) 
solenoid field. Microwave polarization reversals are repeated 
several times while data taking. In order to minimize systematics 
caused by the time dependent variation of the acceptance between 
the microwave reversals, the polarizations are frequently 
reversed by inverting of the solenoid field. 

For the AA studies the double ratios of event numbers, R$_{f}$ , 
is used in the following form:
\begin{equation}
\label{eq4} 
R_f(\phi)=\left[{N_{+,f}^U(\phi)/N_{-,f}^D(\phi)}\right]\cdot 
\left[{N_{+,f}^D(\phi)/N_{-,f}^U(\phi)}\right],
\end{equation}
where $N_{p,f}^t(\phi)$ is a number of events in each $\phi$-bin 
from the target cell $t$, $t=U,\,D$, $p=+$ or $-$ is the sign of 
the target polarization, $f=+$ or $-$ is the direction of the 
target solenoid field. Using Eqs. (\ref{eq1},\ref{eq2},\ref{eq3}) 
with $P_{\pm}$ as an absolute value of averaged products of the 
positive or negative target polarization and dilution factor, the 
number of events can be expressed as
\begin{equation}
\label{eq5}
N_{p,f}^t\!\!=\!C_f^t(\phi)L_{p,f}^t\left[{(B_0\!\!+\!\!
B_1\cos(\phi)\!\!+\!\!B_2\sin(\phi)\!\!+\!
\dots\!)\!\pm\!P_p(A_0\!\!+\!\!A_1\sin(\phi)\!\!+\!\!A_2\sin(2\phi)
\!\!+\!\dots\!)}\right]\,,
\end{equation}
where $C_f^t(\phi)$ is the acceptance factor (source of false 
asymmetries), $L_{p,f}^t$ is a luminosity depending on the beam 
flux and target densities. The coefficients B$_{0}$, B$_{1}$, 
\ldots and $A_{0}$, $A_{1}$, \ldots characterize contributions of 
partial cross sections. Substituting Eq.(\ref{eq5}) in Eq. 
(\ref{eq4}) one can see that the acceptance factors are canceled, 
as well as the luminosity factors if the beam muons cross the 
both cells. So, the ratio $R_{f}(\phi)$ depends only on physics 
characteristics of the SIDIS process and it is expressed via 
asymmetry $a(\phi)$, Eq. (\ref{eq2}), in the quadratic equation, 
approximate solution of which is:
\begin{equation}
\label{eq6}
a_f= \left[{R_f(\phi)-1}\right]/
(P_{+,f}^U+P_{+,f}^D+P_{-,f}^U+P_{-,f}^D )\,.
\end{equation}
Since asymmetry should not depend on the direction of the 
solenoid field, one can expect to have $a_{+}=a_{-}$. Small 
difference between $a_{+}$ and $a_{-}$ could appear due to the 
solenoid field dependent contributions non-factorizable in Eq. 
(\ref{eq5}). But these contributions have different signs and 
canceled in the sum $a(\phi)=a_{+}(\phi)+a_{-}(\phi)$. So, the 
weighted sum $a(\phi)=a_{+}(\phi)+a_{-}(\phi)$, calculated 
separately for each year of data taking and averaged at the end, 
is obtained for the final results.

\paragraph{3. Data selection.}

The data selection, aimed at having a clean sample of hadrons, 
has been performed in three steps, using a preselected sample. 
This sample contained about 167.5M of SIDIS events with $Q^{2}>1$ 
GeV$^{2}$ and $y>0.1$ in a form of reconstructed vertices with 
incoming and outgoing muons and one or more additional outgoing 
tracks. 

1. "GOOD SIDIS EVENTS" have been selected out of the preselected 
ones applying more stringent cuts on the quality of reconstructed 
tracks and vertices, vertex positions inside the target cells, 
momentum of the incoming muon (140-180 GeV/c), energy transfer 
($y<0.9$) and invariant mass of the final states ($W>5$ GeV). About 
58{\%} of events of the initial sample have survived after these 
cuts.

2. "GOOD TRACKS'' (about 157 M) have been selected out of the 
total tracks (about 290 M) from GOOD SIDIS EVENTS excluding the 
tracks identified as muons and tracks with $z>1$ and $p_h^T<0.1$ GeV/c. 

3. "GOOD HADRONS" from GOOD TRACKS have been identified using the 
information from the hadron calorimeters HCAL1 and HCAL2. Each of 
the GOOD TRACKS is considered as the GOOD HADRON if: this track 
hits one of the calorimeter, the calorimeter has the energy 
cluster associated with this hit with $E_{HCAL1}>5$ GeV, or 
$E_{HCAL2}>7$ GeV, coordinates of the cluster are compatible with 
coordinates of the track and the energy of the cluster is 
compatible with the momentum of the track.

The total number of GOOD HADRONS was about 53 M. Each of the GOOD 
HAD\-RONS is included to the asymmetry evaluations. 

\paragraph{4. Results.}

\begin{figure}[h!]
\begin{center}
\includegraphics[width=0.45\textwidth,height=0.31\textwidth]
{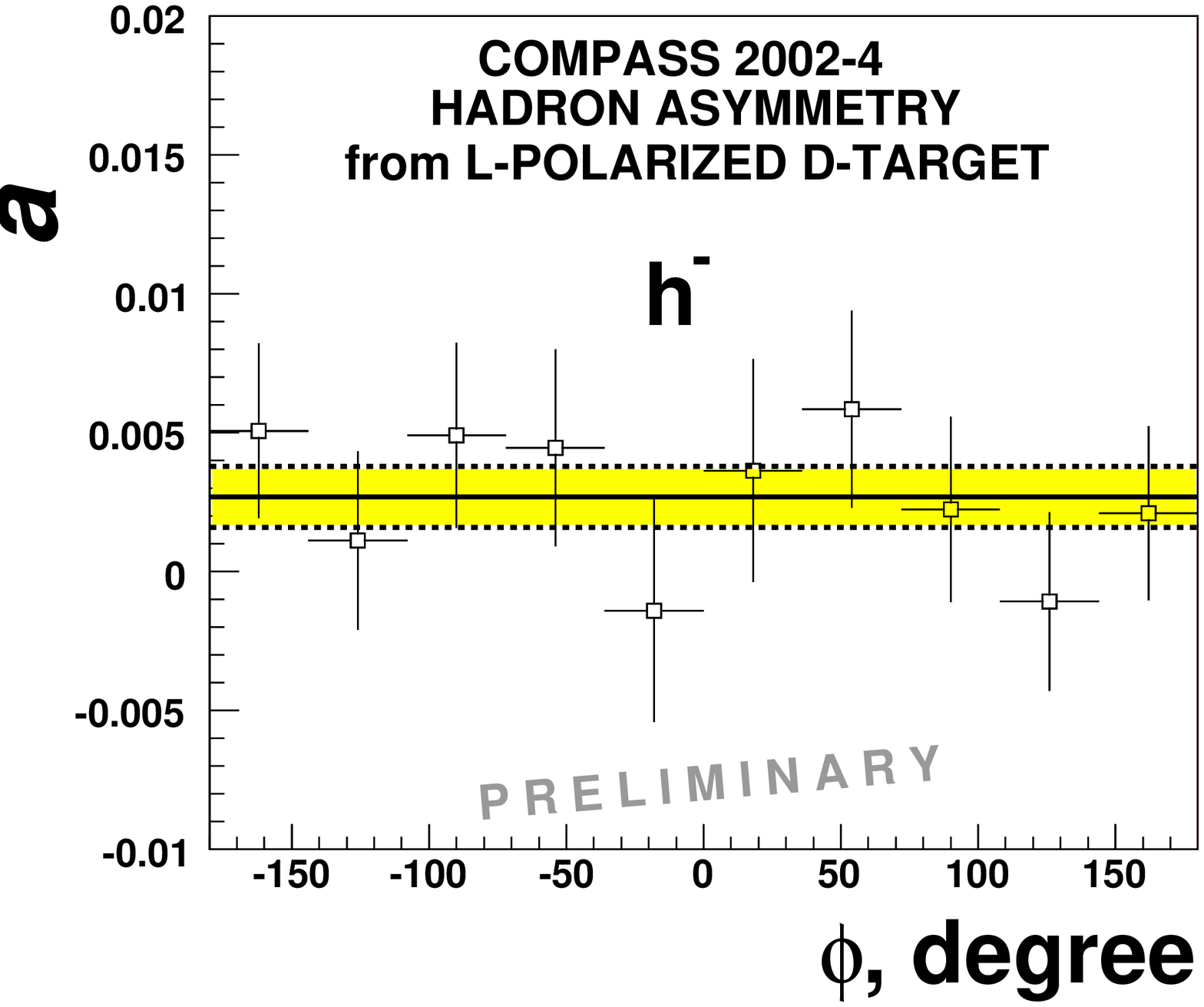}
\includegraphics[width=0.45\textwidth,height=0.31\textwidth]
{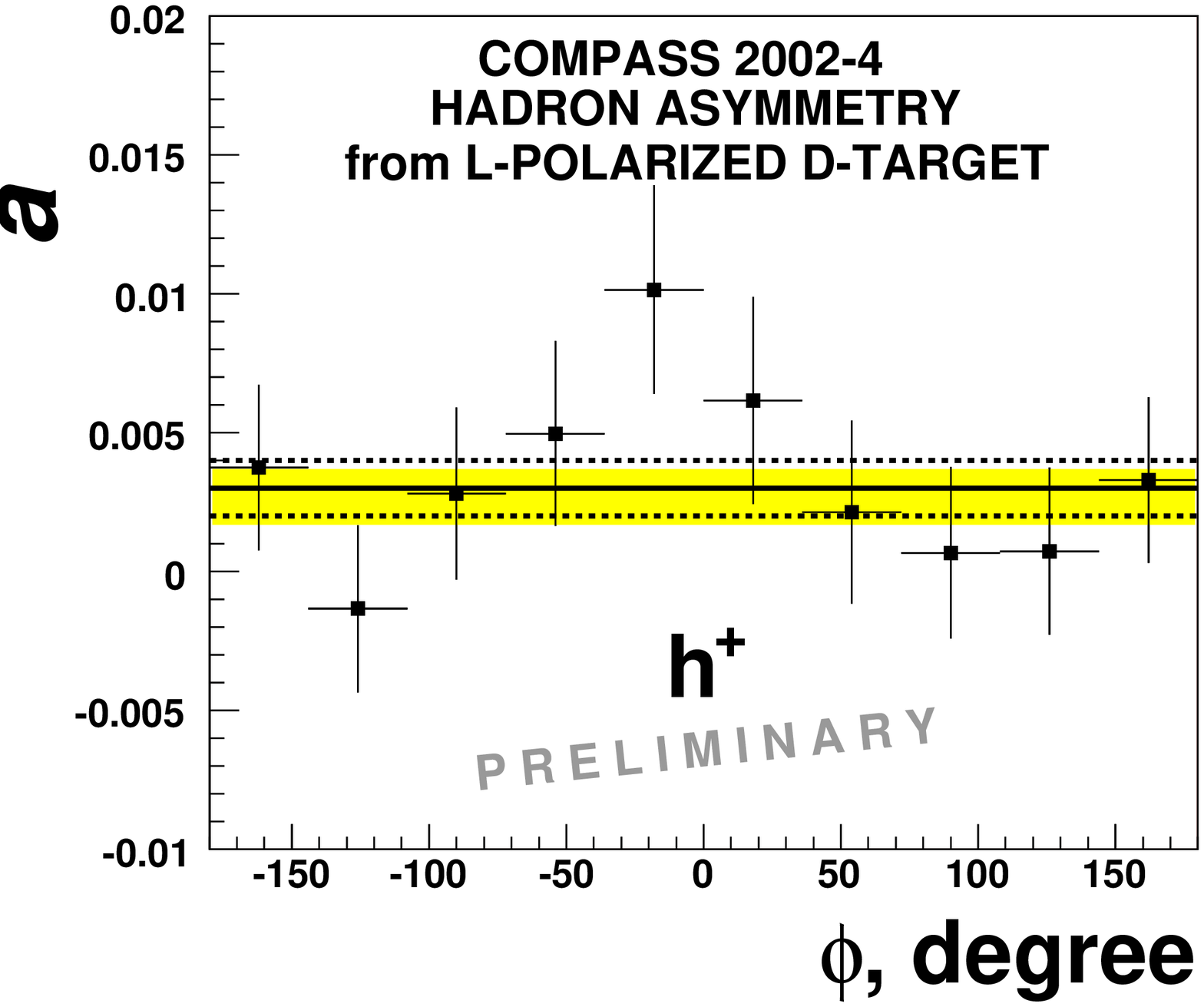}
\end{center}
\vskip-7mm
\caption
{\label{fig2}\footnotesize Azimuthal asymmetries a($\phi )$ for 
negative (left) and positive (right) hadrons and results of fits 
by the constants with $\chi^2/d.f.$ equal to 3.4/5 (5.2/5), 
respectively.}
\end{figure}

The weighted sum of azimuthal asymmetries a($\phi )$=a$_{ + 
}(\phi )$+a$_{ - }(\phi )$, averaged over all kinematical 
variables, are shown in Fig. \ref{fig2} for negative and positive 
hadrons. They have been fitted by functions

\begin{equation}
\label{eq7}
a(\phi)=a^{\rm const}+a^{\sin\phi}\sin(\phi)+a^{\sin2\phi}\sin(2\phi)+ 
a^{\sin3\phi}\sin(3\phi)+a^{\cos\phi}\cos(\phi).
\end{equation}

The fit parameters, characterizing $\phi$-modulation amplitudes, 
are compatible with zero. The $\phi$-independent parts of 
a($\phi)$ differ from zero and are almost equal for $h^{-}$ and 
$h^{+}$. The fits of $a(\phi)$ by constants are also shown is 
Fig. \ref{fig2}.

As already specified, the $\phi$-independent parts of asymmetries 
come from the $d\sigma_{LL}$ contributions to the cross sections, 
which are proportional to helicity PDF times PFF (see Eq. 
(\ref{eq3})) of non-polarized quarks in the non-polarized hadron. 
For the deuteron target this contribution is expected to be 
charge independent.

\begin{figure}[h!]
\vskip-3mm
\begin{center}
\hspace{-5mm}
\includegraphics[width=0.32\textwidth]
{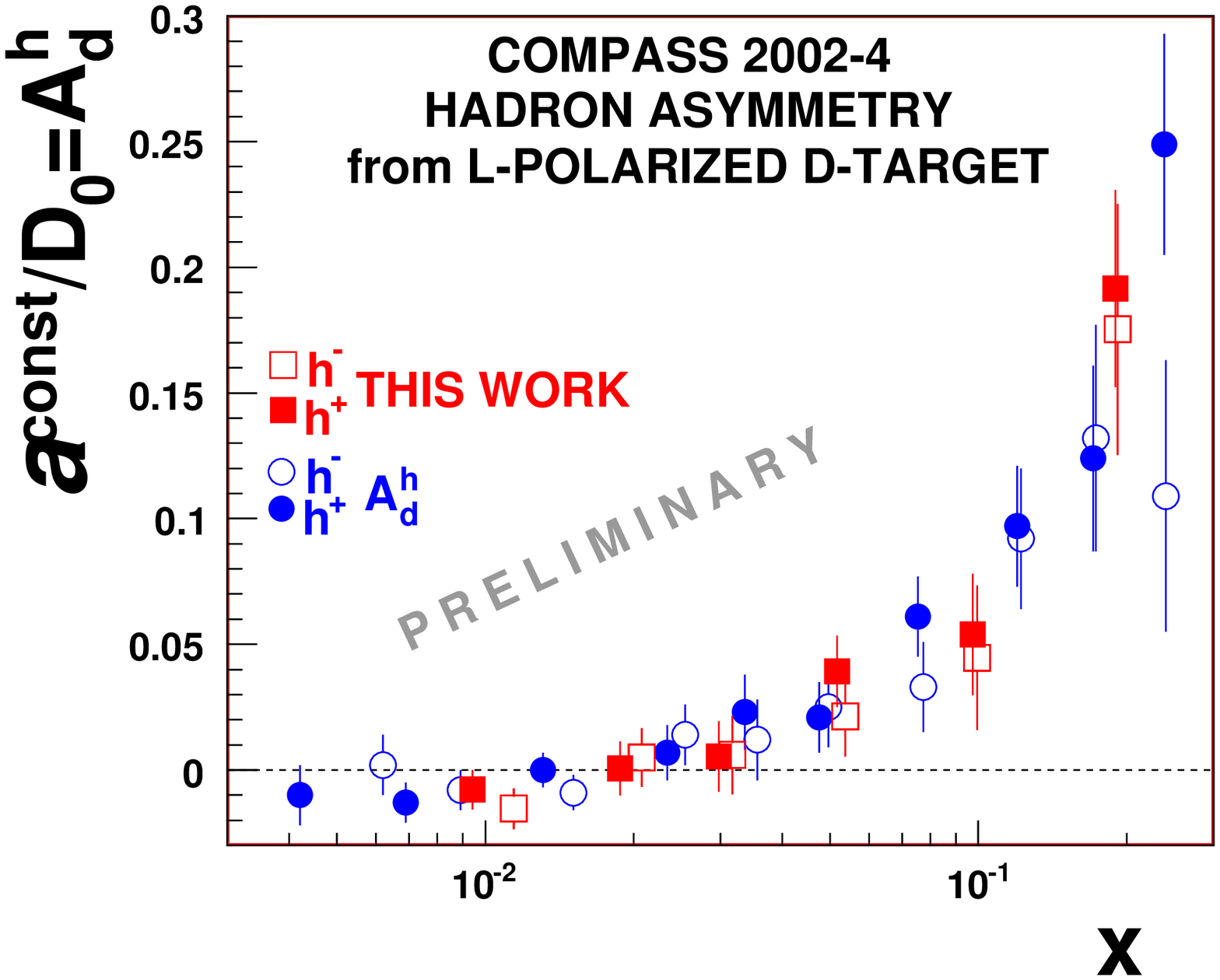}\hspace{-1mm}
\includegraphics[width=0.32\textwidth]
{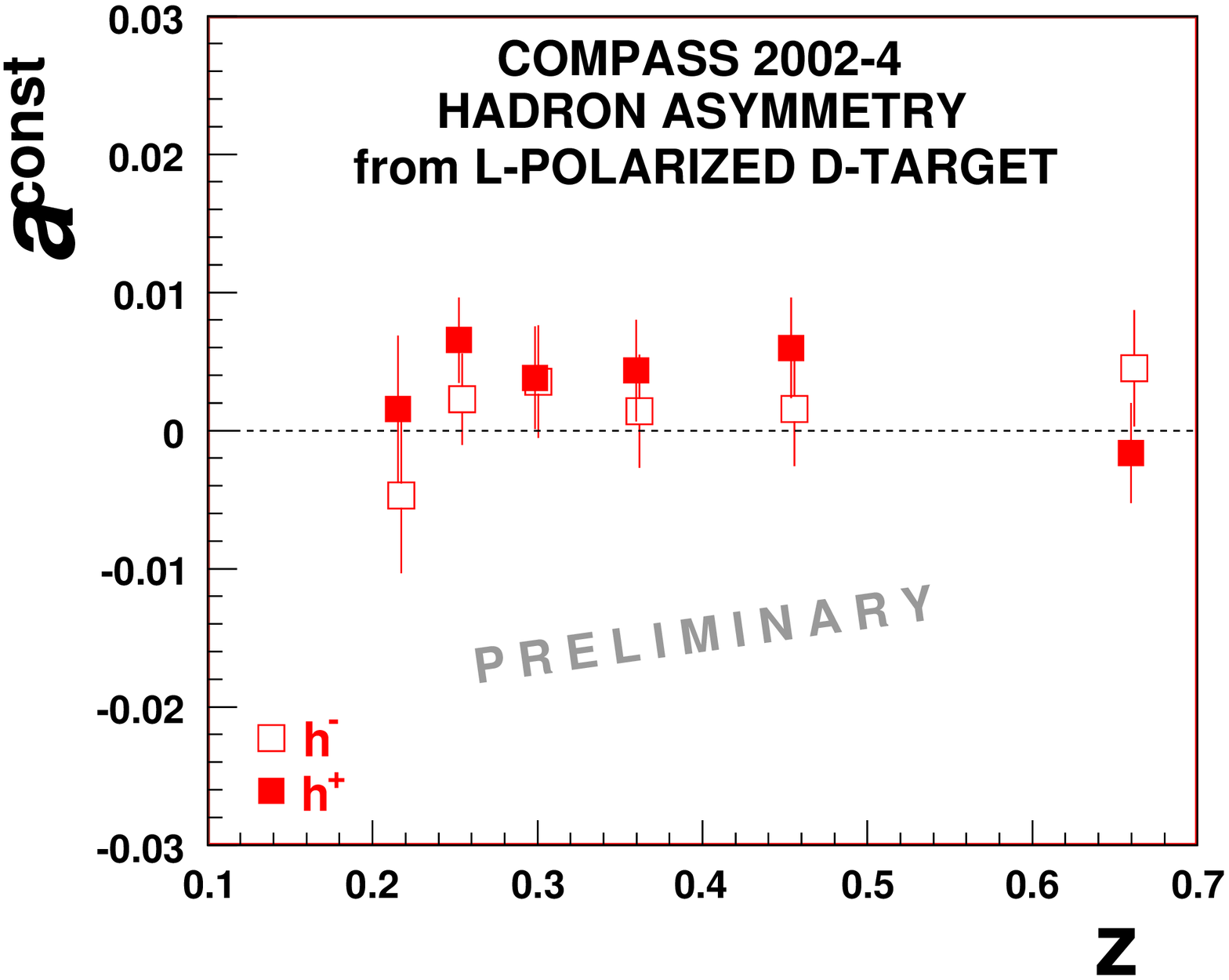}\hspace{-2mm}
\includegraphics[width=0.32\textwidth]
{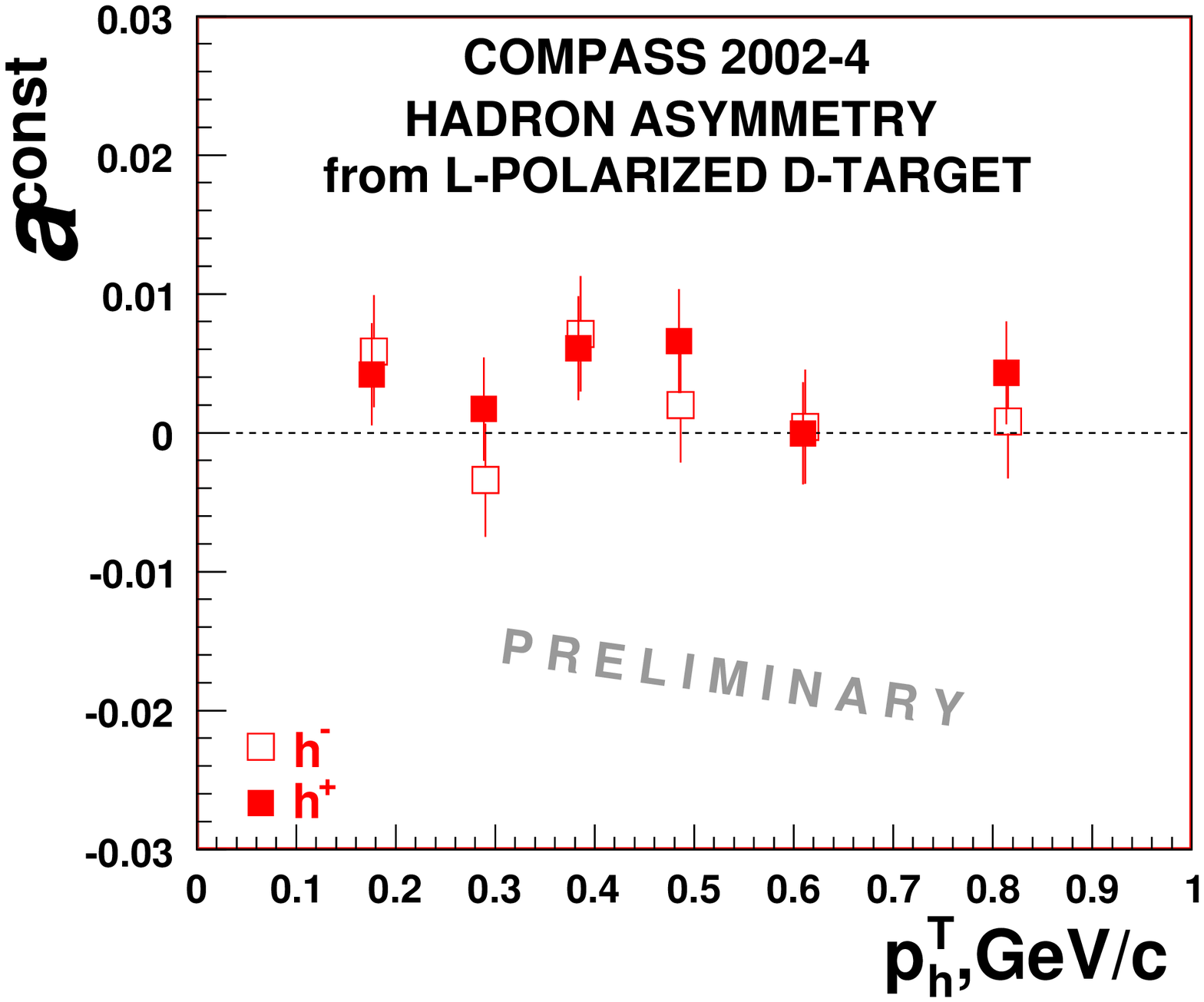}
\end{center}
\vskip-5mm
\caption
{\label{fig3} \footnotesize Dependence of the  AA fit parameters 
$a^{\rm const}$ on kinematical variables.}
\end{figure}

\begin{figure}[hb!]
\begin{center}
\hspace{-3mm}
\includegraphics[width=0.32\textwidth]
{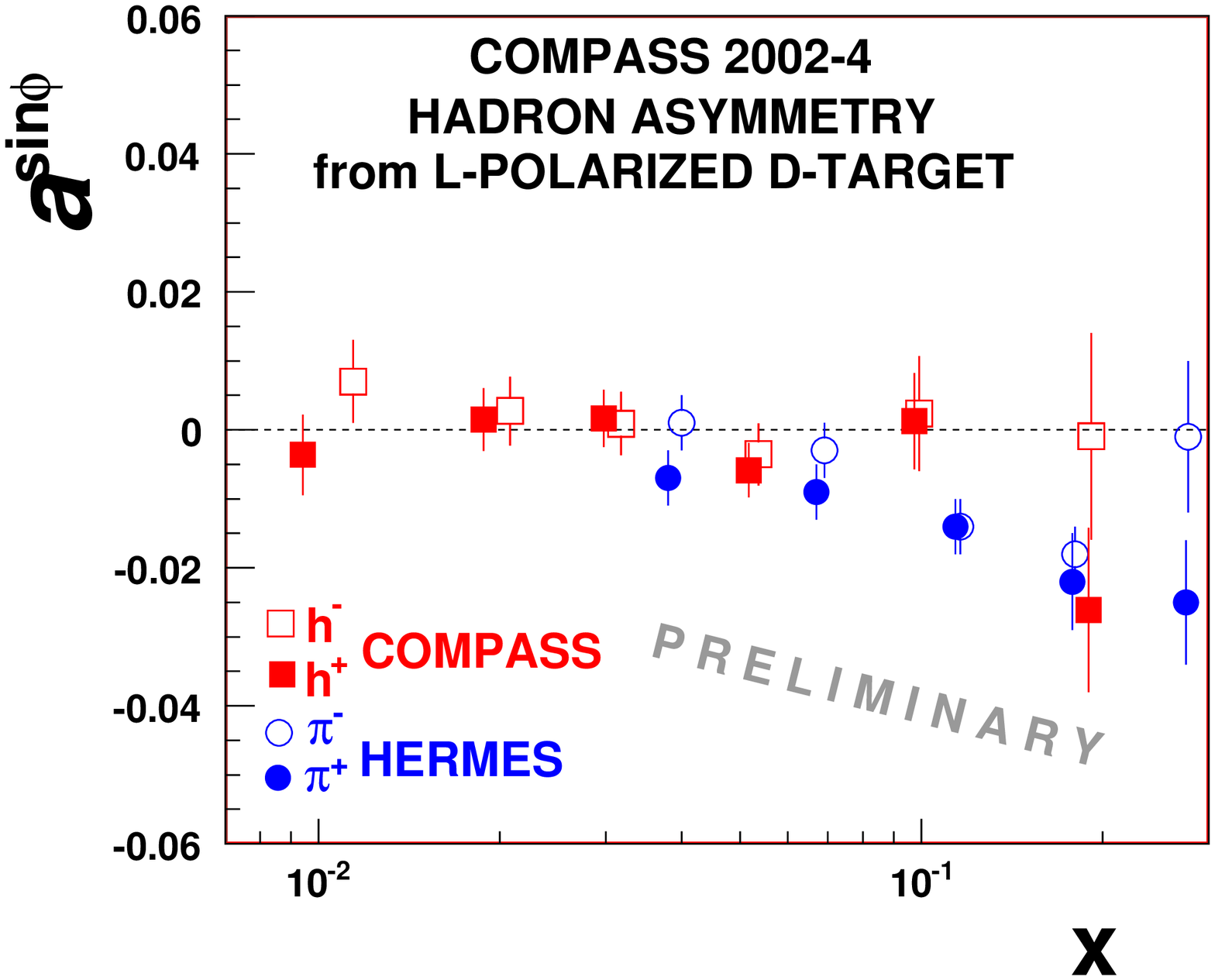}\hspace{-1mm}
\includegraphics[width=0.32\textwidth]
{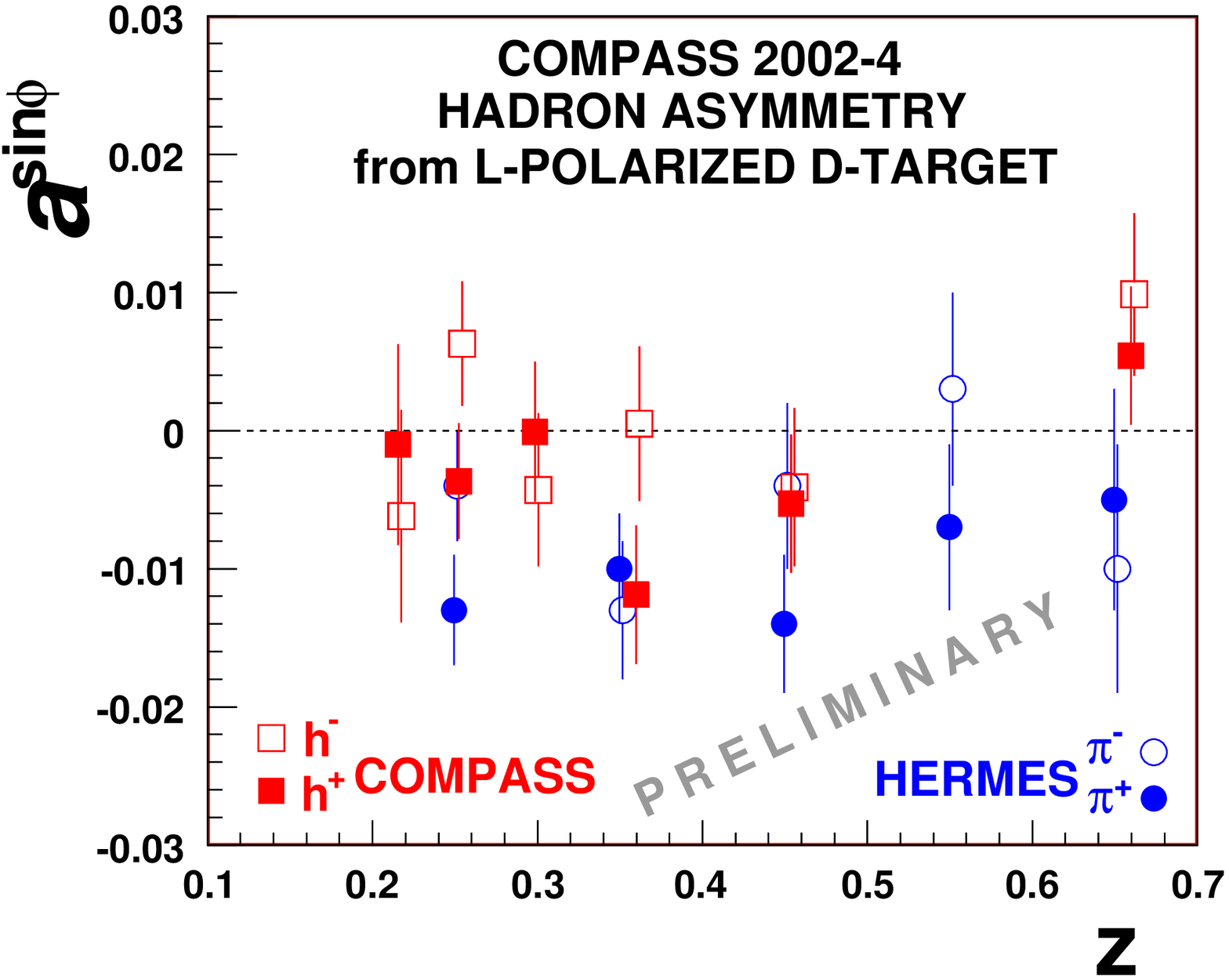}\hspace{-2mm}
\includegraphics[width=0.32\textwidth]
{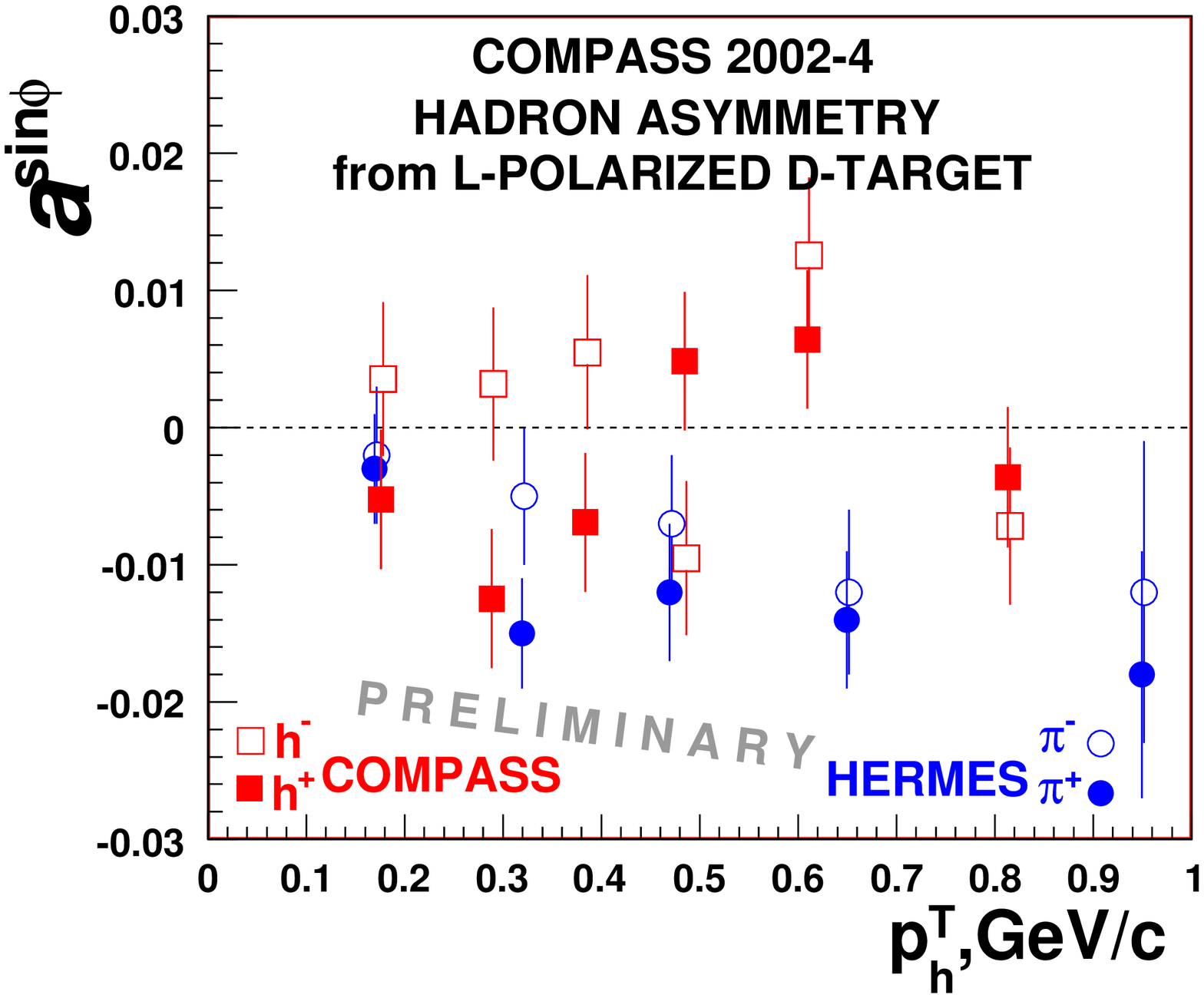}
\end{center}
\vskip-5mm
\caption
{\label{fig4} \footnotesize Dependence of the AA fit parameters 
$a^{\sin\phi}$ on kinematical variables and similar data of HERMES 
\cite{[7]} for identified leading pions.}
\end{figure}

\begin{figure}[h!]
\begin{center}
\hspace{-3mm}
\includegraphics[width=0.32\textwidth]
{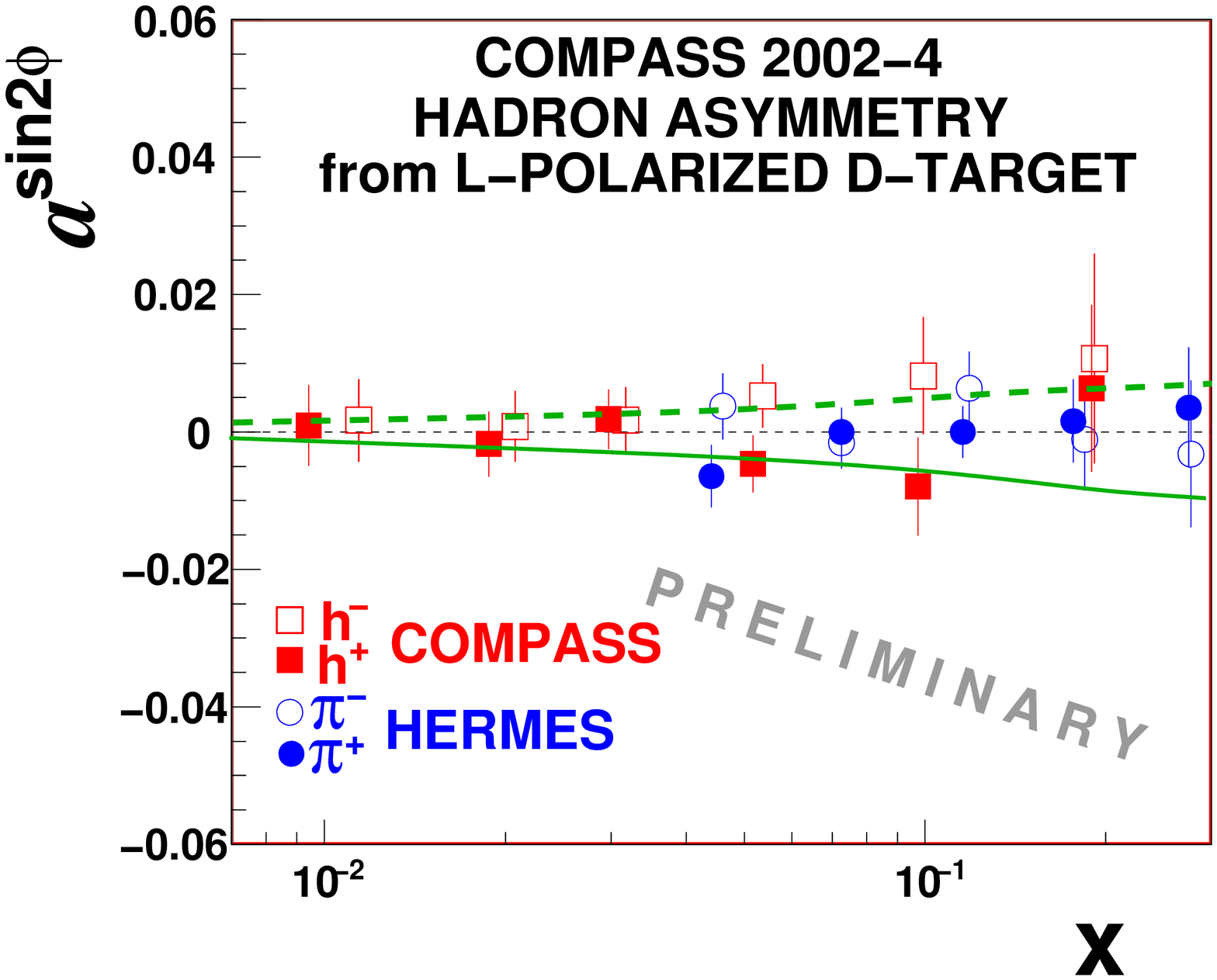}\hspace{-1mm}
\includegraphics[width=0.32\textwidth]
{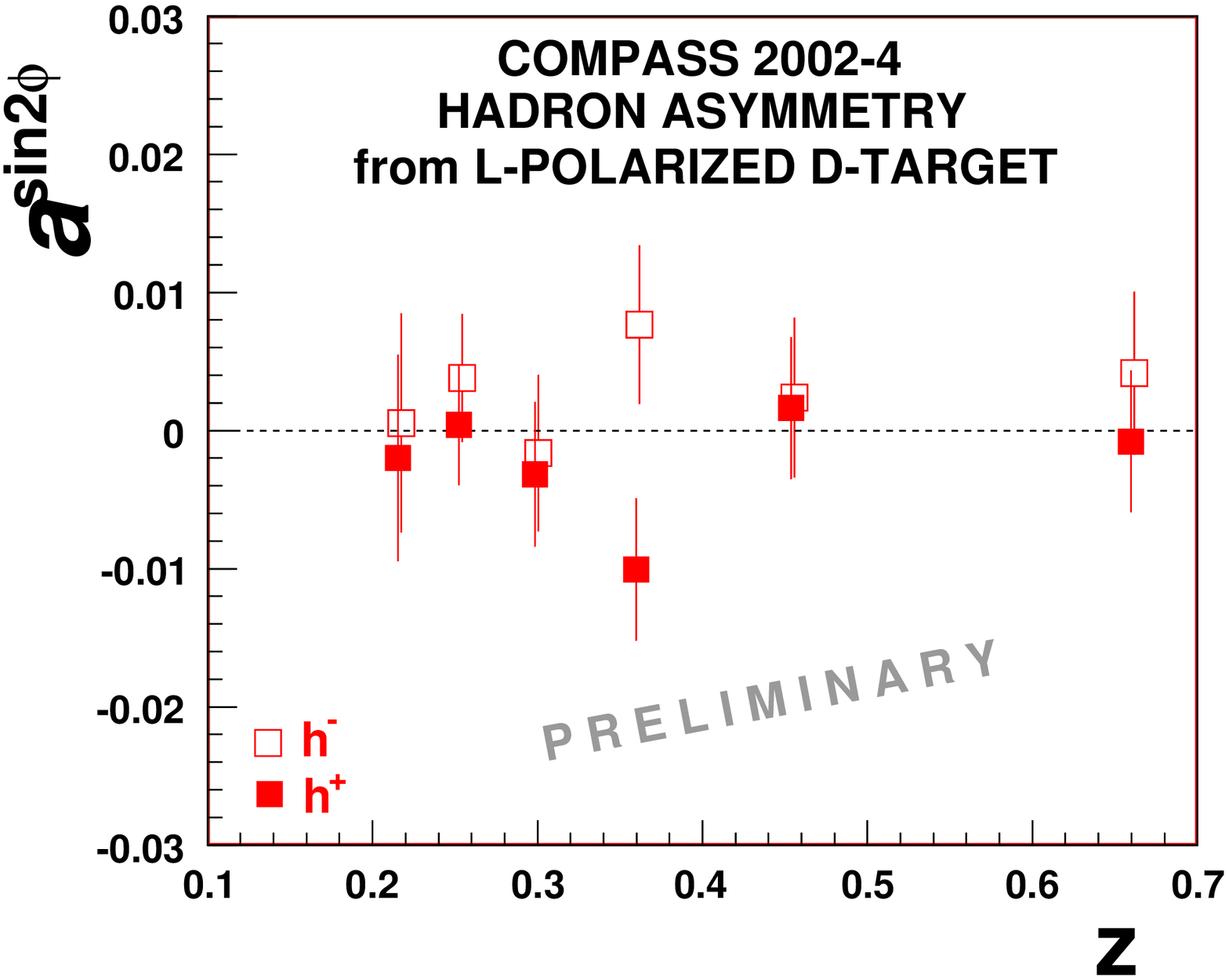}\hspace{-2mm}
\includegraphics[width=0.32\textwidth]
{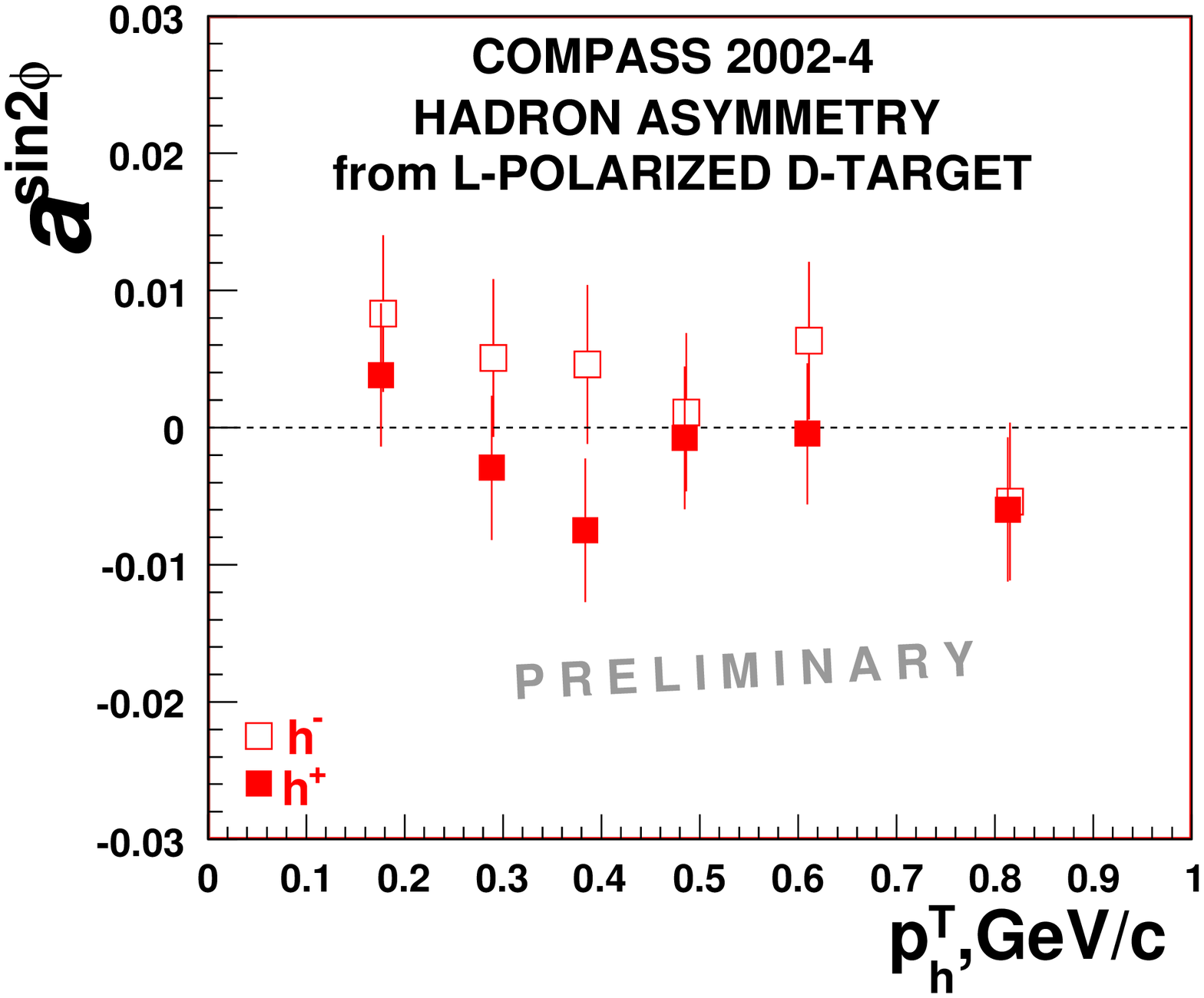}
\end{center}
\vskip-5mm
\caption
{\label{fig5} \footnotesize Dependence of the AA fit parameters 
$a^{\sin2\phi}$ on kinematical variables compared to the data of 
HERMES and calculations by H.Avakian et al. \cite{[13]}: dashed 
line -- $h^{-}$, solid line -- $h^{+}$.}
\end{figure}

\begin{figure}[h!]
\begin{center}
\hspace{-3mm}
\includegraphics[width=0.32\textwidth]
{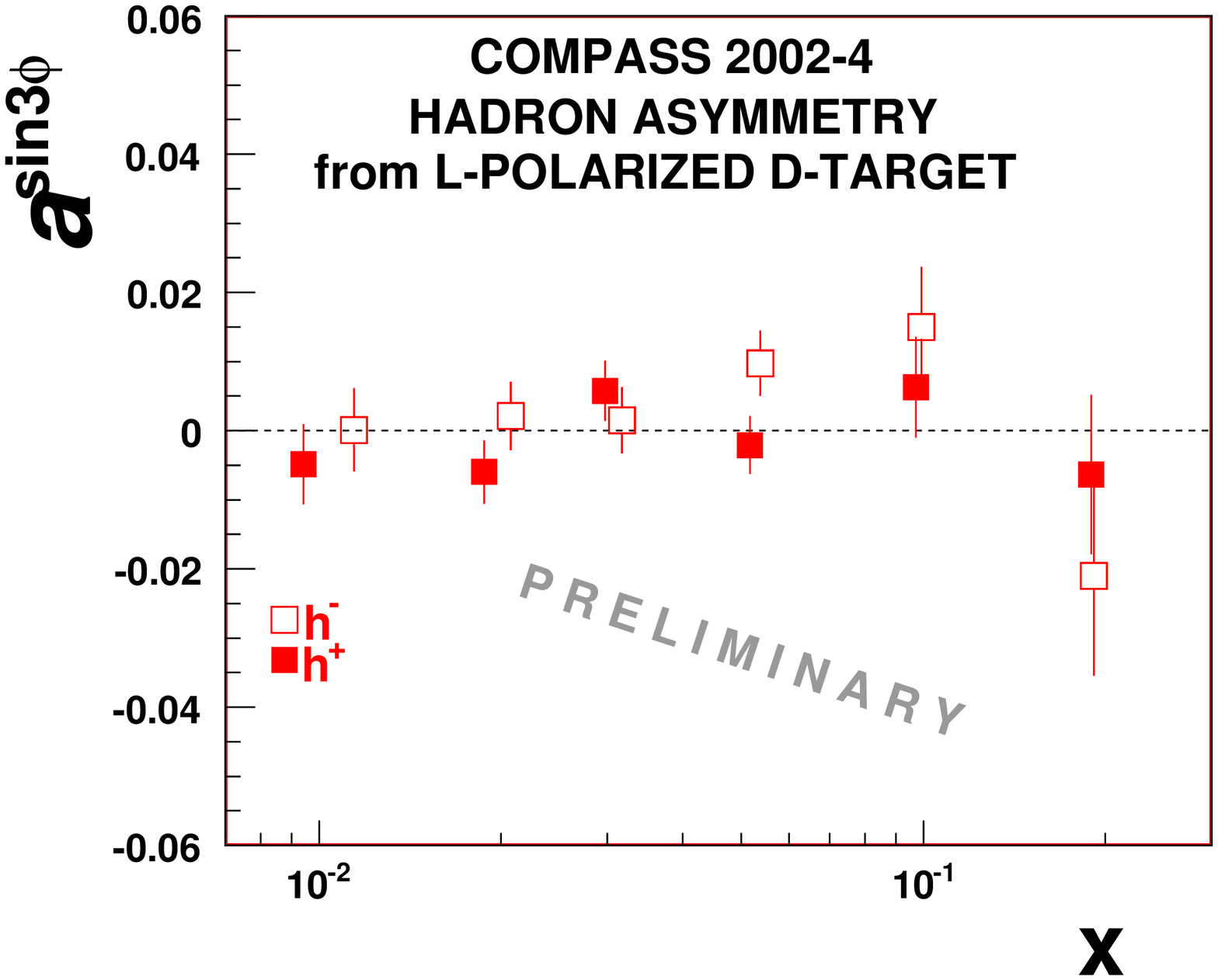}\hspace{-1mm}
\includegraphics[width=0.32\textwidth]
{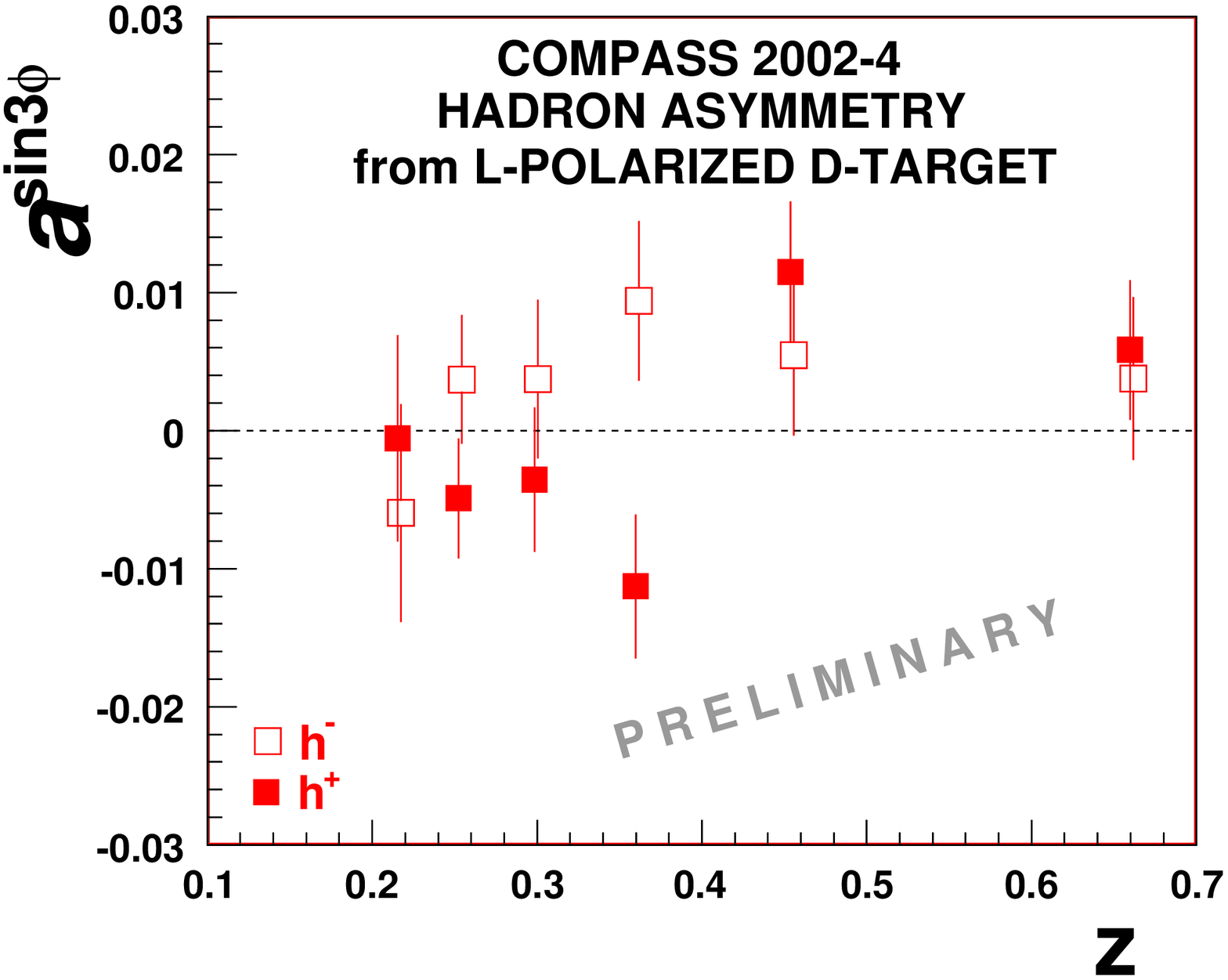}\hspace{-2mm}
\includegraphics[width=0.32\textwidth]
{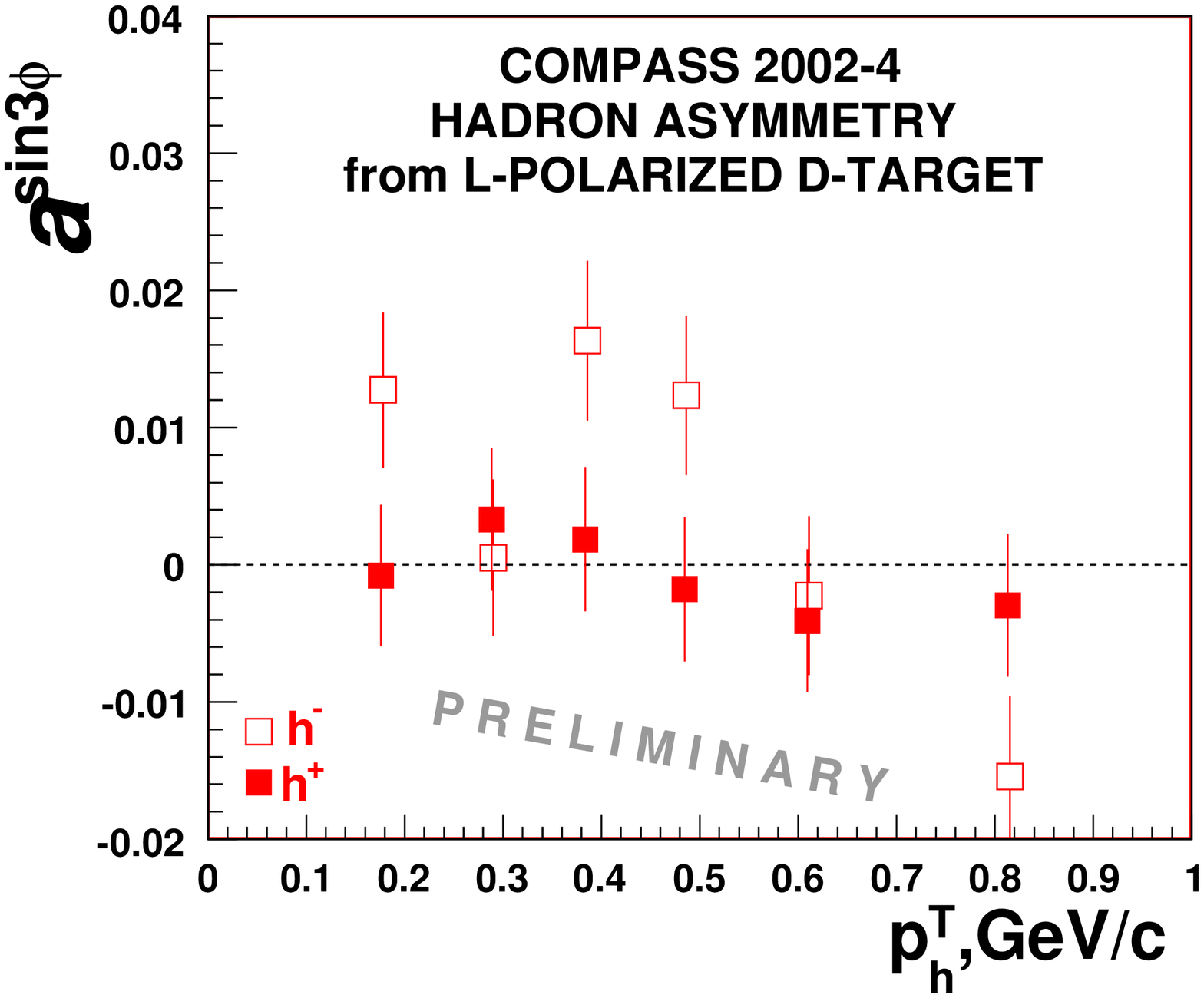}
\end{center}
\vskip-5mm
\caption
{\label{fig6} \footnotesize Dependence of the AA fit parameters 
a$^{\sin3\phi }$ on kinematical variables.}
\end{figure}

\begin{figure}[h!]
\begin{center}
\hspace{-3mm}
\includegraphics[width=0.32\textwidth]
{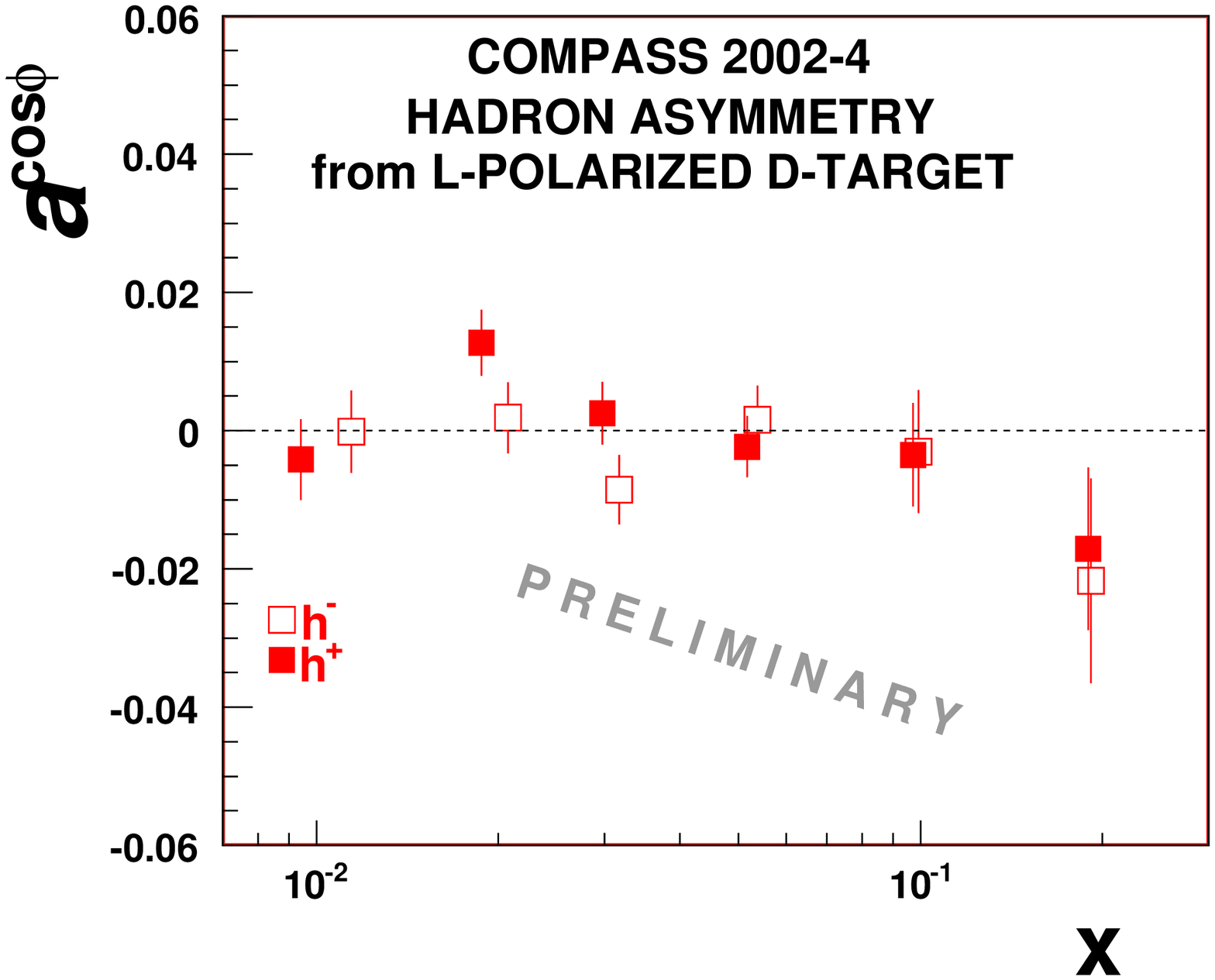}\hspace{-1mm}
\includegraphics[width=0.32\textwidth]
{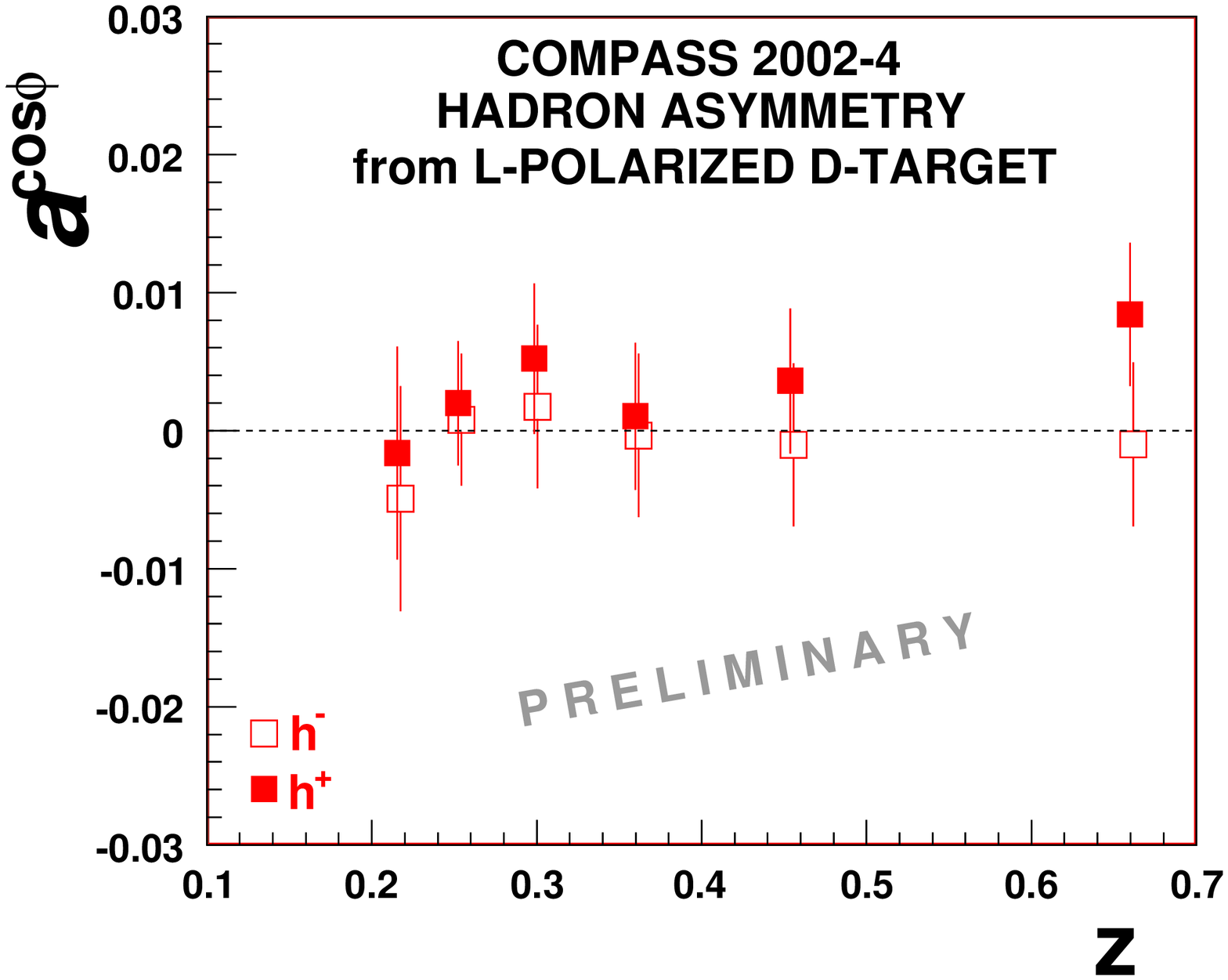}\hspace{-2mm}
\includegraphics[width=0.32\textwidth]
{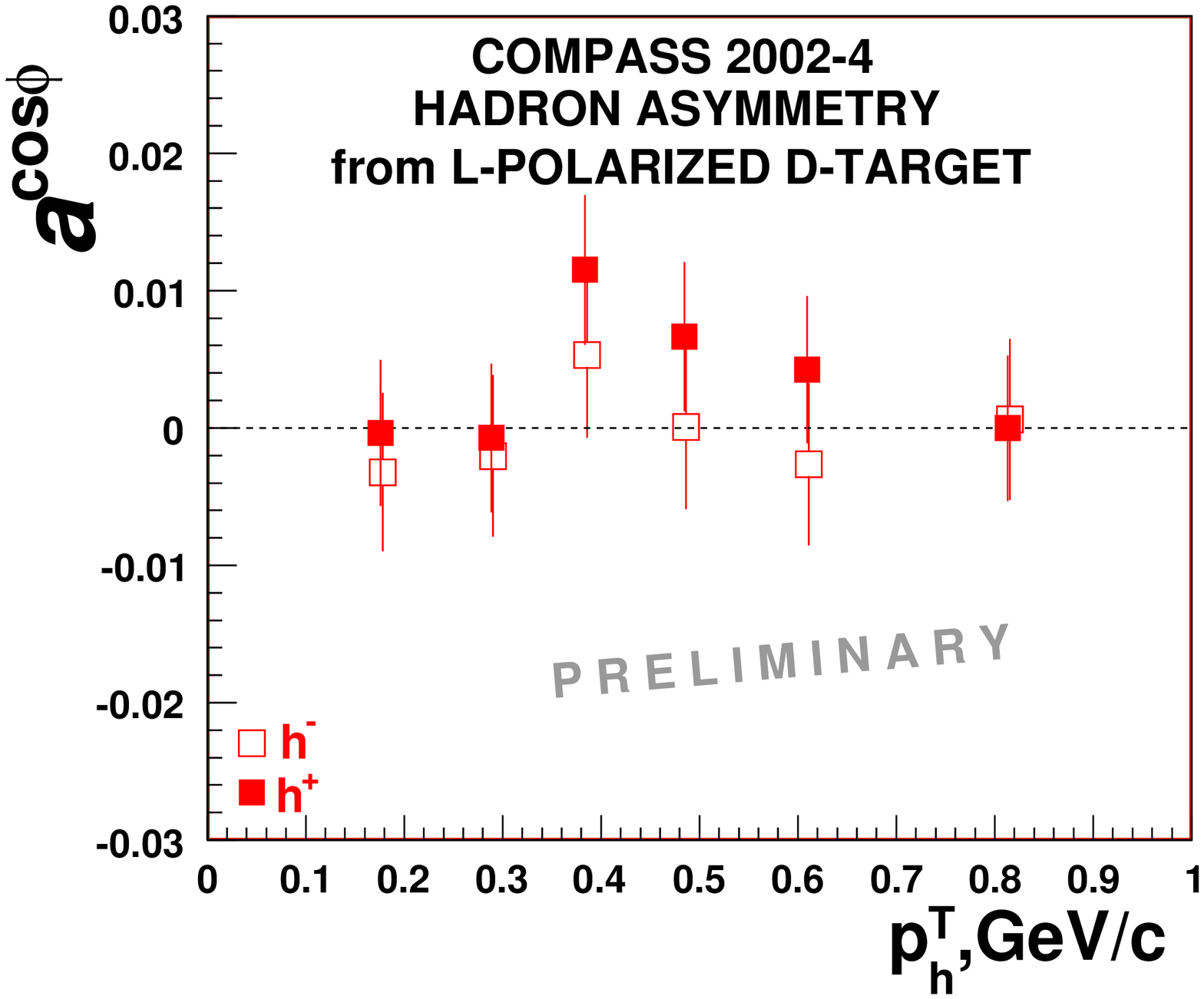}
\end{center}
\vskip-5mm
\caption
{\label{fig7}\footnotesize Dependence of the AA fit parameters 
a$^{cos\phi }$ on kinematical variables.}
\end{figure}

Dependence of the AA fit parameters on the kinematical variables 
are shown in Figs. \ref{fig3}-\ref{fig7}. 

The parameters $a^{\rm const}(x)$, being divided by the virtual 
photon depolarization factor $D_{0}$, are equal (by definition) 
to the asymmetry $A_d^h(x)$, already published by COMPASS 
\cite{[12]}. Agreement of these data and data of the present 
analysis has demonstrated internal consistency of the results.

The x-dependence of the $\sin(\phi)$ modulations of the AA, 
observed by HERMES, is less pronounced at COMPASS. This 
modulation is due to pure twist-3 PDF's entering from the 
$d\sigma_{0L}$ contribution to the AA with a factor $Mx/Q$.

The amplitudes of the $\sin(2\phi)$ modulations are small, 
consistent with zero within the errors. They could be caused by 
PDF $h_{1L}^\bot$ in $d\sigma_{0L}$.

Some peculiarities of the data on the $a^{\sin3\phi}$ are seen 
from Fig. \ref{fig6}, for instance, the points for $h^{-}$ are 
mostly positive while for $h^{+}$ they are mostly negative like 
for the COMPASS results from the transversally polarized target 
\cite{Kotzinian:2007uv}. Remind that this modulation could come 
from the pretzelosity PDF $h_{1T}^\bot$ in $d\sigma _{0T}$, 
additionally suppressed by $\sin(\theta _\gamma)\sim xM/Q$.

The $\cos(\phi)$ modulation of the AA is studied for the first 
time. It is mainly due to a pure twist-3 PDF $g_L^\bot$ in 
$d\sigma_{LL}$, an analog to the Cahn effect \cite{[15]} in 
unpolarized SIDIS.

\paragraph{5. Conclusions and prospects.} 
\begin{enumerate}
\itemsep-1mm
\item 
The azimuthal asymmetries $(a(\phi))$ in the SIDIS ($Q^{2}>1$ 
GeV$^2$, $y>0.1$) production of negative ($h^{-}$) and positive 
($h^{+}$) hadrons by 160 GeV muons on the longitudinally 
polarized deuterium target, have been studied with the COMPASS 
data collected in 2002 -- 2004. 
\item 
For the integrated over $x$, $z$ and $p_h^T$ variables all 
$\phi$-modulation amplitudes of $a(\phi)$ are consistent with 
zero within errors, while the $\phi$-independent parts of the 
$a(\phi)$ differ from zero and are almost equal for $h^{-}$ and 
$h^{+}$.
\item 
The amplitudes as functions of kinematical variables are 
studied in the region of $x=0.004-0.7$, $z=0.2-0.9$, $p_h^T=0.1-1$
GeV/c. It was found that:
\begin{itemize} 
\itemsep-1mm
\item $\phi$-independent parts of the a($\phi)$, 
$a^{\rm const}(x)/D_0=A^h_d$, where D$_{0}$ is a virtual photon 
depolarization factor, are in agreement with the COMPASS 
published data \cite{[12]} on $A^h_d$, calculated by another 
method and using different cuts;

\item the amplitudes $a^{\sin\phi}(x,z,p_h^T)$ are small and 
in general do not contradict to the HERMES data \cite{[7]}, if 
one takes into account the difference in $x$, $Q^{2}$ and $W$ 
between the two experiments. One can also note, that in the 
HERMES experiment the asymmetries are calculated for identified 
leading pions, while in this analysis every hadron is included in 
the asymmetry evaluations;

\item the amplitudes $a^{\sin2\phi}$, $a^{\sin3\phi}$ and 
$a^{\cos\phi}$ are consistent with zero within statistical errors 
of about 0.5{\%} (only statistical errors are shown in the plots 
while systematic errors are estimated to be much smaller).
\end{itemize}
\item The results of this analysis are obtained with restriction 
$z>0.2$ of the energy fraction of the hadron in order to assure 
that it comes from the current fragmentation region. This request 
removes almost one half of statistics. The tests have shown that 
with a lower cut, $z>0.05$, the results are identical. 

\item The reported data are preliminary. New data of 2006 from the 
deuterium target will be added. These data will increase the 
statistics by about a factor of 2. New data of 2007 from the 
hydrogen target will be very interesting in comparison with the 
effects already observed by the COMPASS and HERMES on the 
transversally polarized targets.
\end{enumerate}

\end{document}